# Efficient four-wave mixing at the nanofocus of integrated organic gap plasmon waveguides on silicon

**Michael P. Nielsen, Xingyuan Shi, Paul Dichtl, Stefan A. Maier, and Rupert F. Oulton***
Department of Physics, Imperial College London, London, SW7 2AZ, UK
*r.oulton@imperial.ac.uk

**Nonlinear optics, especially frequency mixing, underpins modern optical technology and scientific exploration in quantum optics[1,2], materials and life sciences[3–5], and optical communications[6–8]. Since nonlinear effects are weak, efficient frequency mixing must accumulate over large interaction lengths restricting the integration of nonlinear photonics with electronics and establishing limitations on mixing processes due to the requirement of phase matching[9]. In this work, we report efficient four-wave mixing (FWM) over micron-scale interaction lengths at telecoms wavelengths. We use an integrated plasmonic gap waveguide on silicon[10,11] that strongly confines light within a nonlinear organic polymer in the gap. Our approach is so effective because the gap waveguide intensifies light by efficiently nanofocusing[10,12] it to a mode cross-section of a few tens of nanometres, generating a nonlinear response so strong that efficient FWM accumulates in just a micron. Our device's capability to rapidly nanofocus and nano-defocus ensures low insertion loss. This is significant as our technique opens up nonlinear optics to a regime where phase matching and dispersion considerations are relaxed, giving rise to the possibility of compact, broadband, and efficient frequency mixing on a platform that can be integrated with silicon photonics[10].**

Four-wave mixing (FWM) is an important nonlinear frequency conversion technique used in photonic integrated circuits and optical communication networks for signal regeneration[8], switching[13], phase-sensitive amplification[14], metrology[15], and entangled photon-pair generation[16]. As a third order nonlinear effect, FWM is extremely sensitive to enhancement by the optical confinement of nanoplasmonic systems[10]. For example, FWM has been demonstrated in a variety of metallic nanostructures including nano-antennas[17], rough surfaces[18,19], and at sharp tips[20]. Nonetheless, efficient frequency conversion has remained elusive. While metals can be highly nonlinear and afford extreme optical localization, at telecommunications wavelengths only a small fraction of a plasmonic mode interacts with the metal and increasing this only exacerbates absorption. An alternative strategy is to incorporate low-loss nonlinear materials within nanoplasmonic systems[21,22]. Indeed, recent theoretical studies of FWM in plasmonic waveguides incorporating nonlinear polymers are promising[23]. Nonlinear polymers defy Miller's rule by exhibiting large Kerr indices[24] for relatively low refractive indices, and this has been exploited in recent studies[25,26]. In the context of plasmonics, this brings two advantages: polymers are straightforward to integrate within metallic nanostructures by solution processing[27], and their low refractive index minimizes propagation loss.

In this letter we utilize a silicon hybrid gap plasmon waveguide (HGPW)[10,11,23] to mediate pump degenerate four-wave mixing (DFWM) in the nonlinear polymer poly[2-methoxy-5-(2-ethylhexyloxy)-1,4-phenylenevinylene] (MEH-PPV)[24], as illustrated in Figure 1 (see Methods). The device consists of input and output gratings to launch and collect optical signals, either side of a metallic waveguide of length, $L$, and width, $W$, as narrow as $W = 25$ nm, which is accessed via two tapered sections. In recent work[10], we demonstrated this system's capability to enhance more than 100 fold the intensity of light within the narrow waveguide; a process known as adiabatic nanofocusing[12]. In this work, the nonlinear polymer infiltrates the narrow waveguide section, where the optical field is expected to be maximal.

Unlike conventional DFWM, our approach does not require zero-dispersion wavelengths for phase-matching (generally plasmonic waveguides do not exhibit such critical points[23]). Here, phase-matching is irrelevant because the propagation distance is considerably shorter than the DFWM coherence lengths under investigation. We study DFWM near a wavelength of $\lambda = 1500$ nm over a pump to signal bandwidth of $\Delta\lambda = 30$ nm, which for a $W = 25$ nm waveguide has a coherence length of 228 μm, far in excess of the 2 μm propagation length. Even a bandwidth of $\Delta\lambda = 300$ nm near 1500 nm would have a coherence length greater than the propagation length.

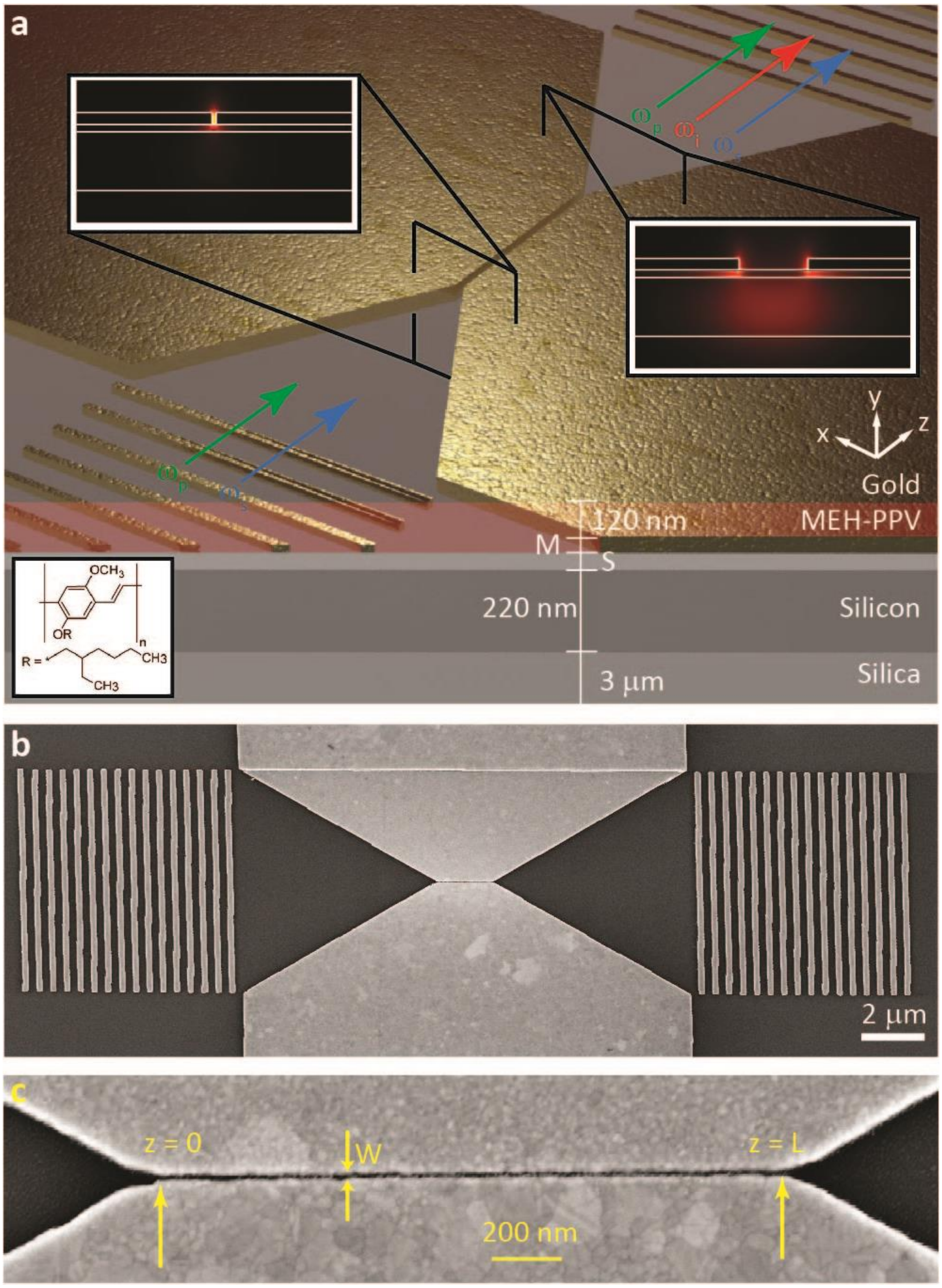

**Figure 1│Nanofocussing devices and modal properties of organic hybrid gap plasmon waveguides on silicon.** (a) Schematic representation of the organic hybrid gap plasmon waveguide. Insets show the nanofocusing mechanism with electromagnetic mode distributions for silicon hybrid gap plasmon waveguide with $S = 25$ nm and $M = 40$ nm for a narrow gap width of $W = 25$ nm and a wide gap width of $W = 500$ nm (the $W = 500$ nm mode has had the field strength increased by a factor of 5 for clarity). Included also is the chemical formula for the nonlinear polymer MEH-PPV. (b) SEM of a $L = 2$ µm long, $W = 25$ nm wide hybrid gap plasmon waveguide without cladding layer depicting the in-/out-coupling gratings. (c) Close-up of the same waveguide.

Figure 1 illustrates the nanofocusing mechanism[10,11]. An input beam polarized parallel to the gratings couples to transverse electric (TE)-like waveguide modes, with dominant electric field component in the plane. For wide gap widths the fundamental TE-like mode propagates primarily in the silicon layer over distances >100 µm as its modal overlap with metal is minimal. For $W < 50$ nm, the mode becomes concentrated in the gap region. While the mode only propagates for a few microns in this state, the gap's field enhancement is dramatic[10]. The taper angle to access this confined mode is selected to minimize propagation loss and reflections or scattering that would reduce the nanofocusing efficiency. For more details on the taper/grating coupling efficiencies and the waveguide propagation losses, the reader is directed to the Supplementary Information.

In order to investigate DFWM in this plasmonic device, two spectrally distinct pulses centered at $\lambda_s = 1450$ nm (signal pulse) and $\lambda_p = 1480$ nm (pump pulse) were generated by filtering a femtosecond pulse centred at $\lambda = 1480$ nm. The spectral full width at half maxima of these pulses were used to estimate transform limited pulse widths of $\geq 1.04$ ps. The pump and signal pulses were coupled to HGPWs through the in-coupling grating and the resulting idler pulse centred at $\lambda_i = \left(2\lambda_p^{-1} - \lambda_s^{-1}\right)^{-1} = 1510$ nm was measured on a spectrometer from the out-coupling grating after spectrally filtering out the pump and signal (see Supplementary Information). Figure 2a compares the normalized input and filtered output spectra for a peak pump power of 30 W and a HGPW with $W = 25$ nm and $L = 2$ µm. The input and output spectral counts represent the power spectral density (PSD, $P(\lambda, z)$) at the start and end of the narrow section of HGPW, respectively, determined from measured grating and tapering efficiencies (see Supplementary Information). Our experimental results agree with theoretical predictions from numerical pulse propagation simulations based on the nonlinear Schrödinger equation (NLSE), using nonlinear material parameters for MEH-PPV, silica and silicon measured with the Z-scan method (see Supplementary Information). Figure 2b shows the simulated conversion spectrum overlaid with experimental pump/signal and idler spectra from Figure 2a.

From the input and output spectra, we can extract the conversion efficiency ($\mathrm{CE}$) of the DFWM process[8], defined as the ratio of the peak idler PSD after the narrow waveguide section ($P(\lambda_i, L)$) to the peak signal PSD at the start ($P(\lambda_s, 0)$) directly from Figure 2a (see Supplementary Information). For this HGPW, $\mathrm{CE} = -13.3$ dB. This was the highest conversion efficiency measured in this study and is comparable to ultrafast DFWM in silicon waveguides over millimeter scale interaction lengths[23]. When considering the integrated PSD of each beam, the generated peak power in the idler is 22.7 % of the power in the signal, assuming identical temporal characteristics of the signal and idler. Accounting for the power used in all beams, we also define a DFWM efficiency from the peak idler power after the plasmonic waveguide, $\mathcal{P}_i(L)$, to both the peak signal power, $\mathcal{P}_s(0)$, and the peak pump power, $\mathcal{P}_p(0)$, at the start of the waveguide, $\eta = \mathcal{P}_i(L)/(\mathcal{P}_s(0)\mathcal{P}_p(0)^2)$. For the most efficient device considered here, $\eta = 0.025\ \%W^{-2}$.

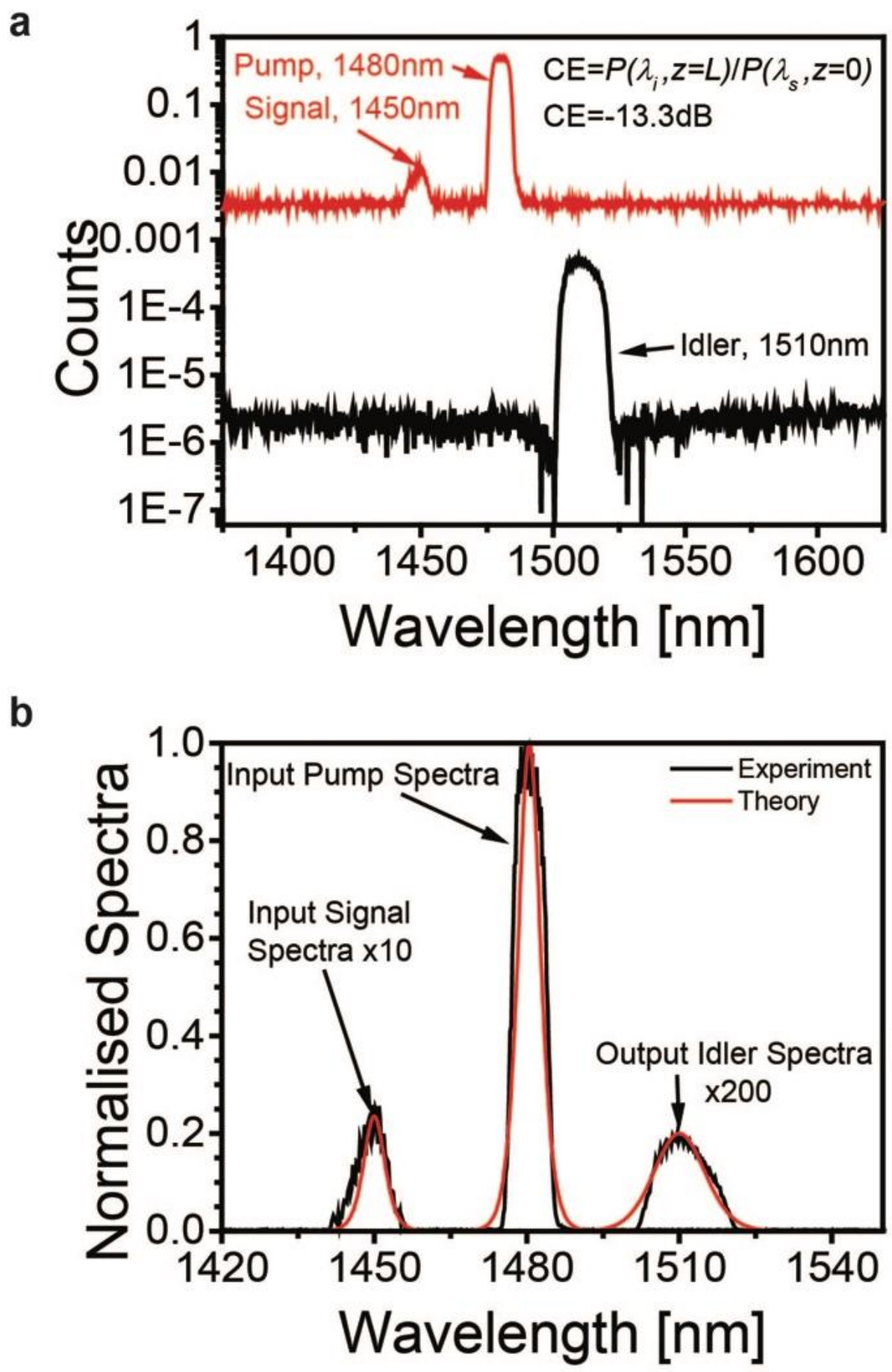

**Figure 2│Four-wave mixing spectra.** (a) Normalised pump/signal spectrum and out-coupled idler spectrum from a $L = 2$ µm waveguide of $W = 25$ nm for $\mathcal{P}_p(0) = 30$ W at repetition rate of 10 kHz. (b) The same spectra (black lines) overlaid by input/output spectra from the NLSE simulations (red lines).

We have confirmed the nature of the conversion process by observing a cubic dependence of idler peak PSD on the combined pump and signal powers (Figure 3a). The cubic relationship arises from the linear dependence on signal power and quadratic dependence on pump power. Figure 3b compares the measured conversion efficiency with that simulated using the NLSE as a function of peak pump power for a HGPW of $W = 25$ nm and $L = 2$ µm. The simulated conversion efficiency varies with peak power cubed until a critical power where nonlinear absorption of the pump beam dominates. For $\mathcal{P}_p(0) > 30$ W in the narrowest waveguides ($W = 25$ nm), the MEH-PPV degraded, setting the upper power limit for our dataset. Although the conversion efficiency roll-off was not observed in experiments, it is remarkable that nonlinear absorption should not limit performance until peak powers approaching 100 W, due to the short device lengths.

The observed conversion efficiencies were approximately 20 dB less than those expected from NLSE simulations using the measured nonlinear parameters of MEH-PPV films. Since the discrepancy was systematic across all measured devices (Figure 4), we can identify a number of reasons. Firstly, poor infiltration of MEH-PPV into the gap would not only affect the waveguide's nonlinear coefficient but also the mode confinment. Secondly, the morphology of the MEH-PPV polymer within the gap could be distinct from that of bulk films, which were used to assess the material's nonlinear parameters with the Z-scan method. Finally, calculating the waveguide nonlinearity, $\gamma$, from the nonlinear responses of the various device materials could require more rigourious theoretical treatment[28]. Nevertheless, all data broadly agrees with theory for a waveguide nonlinearity, $\gamma$, that is a factor of $2.5 - 3$ times less than that inferred from Z-scan measurements.

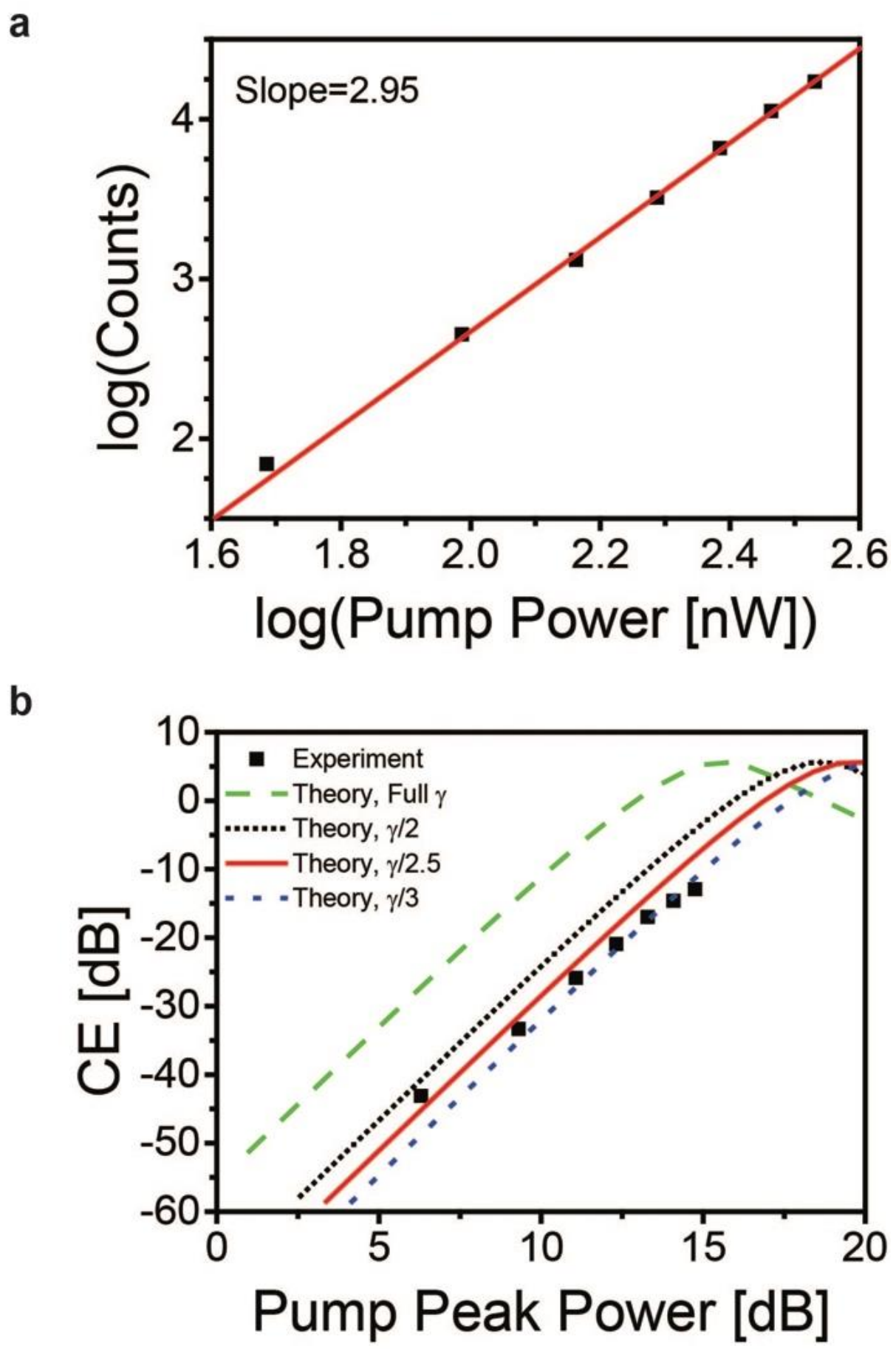

**Figure 3│Degenerate four-wave mixing as a function of combined pump and signal power for a silicon hybrid gap plasmon waveguide ($W = 25$ nm and $L = 2$ μm).** (a) Peak idler photon counts as a function of average pump power showing the third order dependence on the combined pump and signal power. (b) Signal to idler conversion efficiency as a function of peak pump power compared to numerical simulations based on solutions of the nonlinear Schrodinger equation for several values of the waveguide nonlinearity, $\gamma$. For this HGPW, the waveguide nonlinearity $\gamma/2.5 = (3.09 + 0.07i) \times 10^4$ $W^{-1}m^{-1}$.

The plasmonic waveguide width and length clearly influence the DFWM conversion efficiency. While a narrower gap boosts the effective nonlinear coefficient the additional propagation loss limits idler generation. This raises the question: what is the optimal interaction length? Figure 4a shows the conversion efficiency of HGPWs with $W = 25$ nm and $\mathcal{P}_p(0) = 30$ W for $L = 1$ to $5$ μm. The experiment is broadly consistent with the theory that conversion efficiency increases with device length until a maximum is reached due to growing propagation loss. The fact that the conversion efficiency is maximal near the measured propagation length of $1.9 \pm 0.6$ μm suggests that DFWM accumulates rapidly and that the optimal gap width is <25 nm. DFWM for gaps where $W < 25$ nm would be more efficient and accumulate over less than a micron and this will be investigated in further work. The dominant role of confinement in these devices is apparent from the much smaller CEs of HGPWs with $W = 50$ nm despite the increase in peak interaction length (Figure 4b). Figures 4c and 4d show complimentary data on how the conversion efficiency varies with gap width for two fixed HGPW legnths of $L = 3$ μm and $L = 5$ μm, at $\mathcal{P}_p(0) = 30$ W. Although the gap width affects both propagation loss and nonlinear coefficient, broad agreement between NLSE simulations and experiments remains, demonstrating that this frequency mixing approach is robust and repeatable.

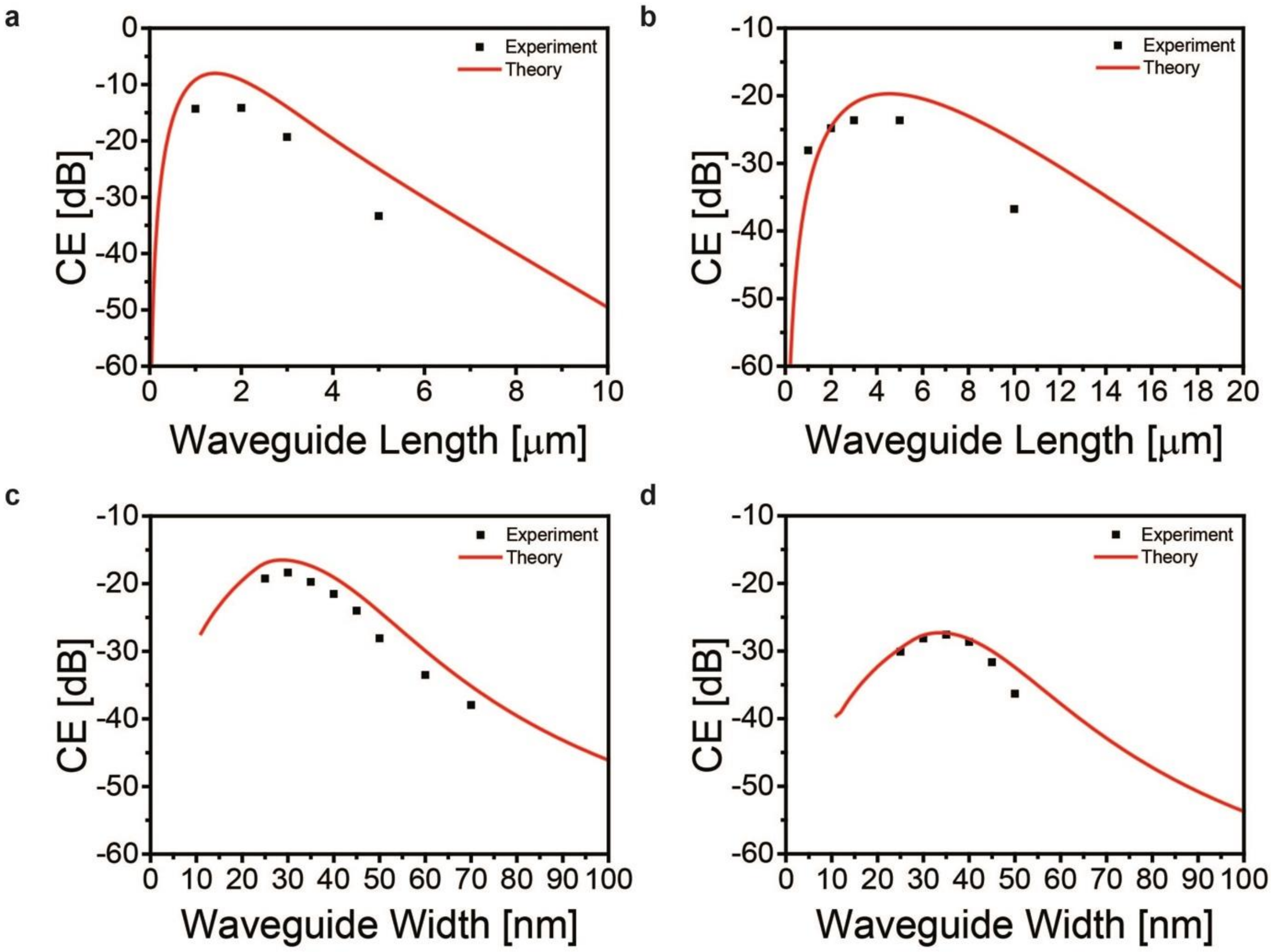

**Figure 4│Degenerate four-wave mixing conversion efficiencies for a variety of different HGPW devices.** Conversion efficiency as a function of waveguide length for HGPWs of (a) $W = 25$ nm at $\mathcal{P}_p(0) = 30$ W and (b) $W = 50$ nm at $\mathcal{P}_p(0) = 40$ W. Conversion efficiency as a function of waveguide width for HGPWs of (c) $L = 3$ μm and (d) $L = 5$ μm at $\mathcal{P}_p(0) = 30$ W. Theoretical conversion efficiencies calculated with the NLSE using a $\gamma$ determined from Z-scan measurements (see Supplementary Infomration) divided by 2.5.

This work proves that plasmonic nanofocusing can be a powerful tool in nonlinear optics. Here we have shown that the intense light within a nanofocus enables nonlinear optical control over extremely short interaction lengths comparable to the vaccuum wavelength of light. Remarkably, at the minimum gap width of 25 nm in this study, we are still operating far from where non-local and quantum effects arise at the sub-nanometre scale[29] suggesting excellent scope for improvement. By achieving an efficient nonlinear frequency mixing process over distances shorter than a plasmonic mode's propagation length, we can eliminate the key problem of insertion loss that has plagued the application of plasmonic systems. Moreover, our approach brings new opportunities to mitigate the limitations introduced by phase matching. By enabling broadband frequency conversion without requiring dispersion engineering or phase matching considerations with considerable room for improvement and elaboration, this platform opens up a new avenue for research combining nanoscale plasmonic devices and nonlinear optics.

## Acknowledgements

This work was sponsored by the UK Engineering and Physical Sciences Research Council (EP/I004343/1 & EP/M013812/1) and Natural Sciences and Engineering Research Council of Canada (NSERC). M.P.N was supported by an NSERC scholarship and EPSRC studentship. S. A. M. acknowledges the Lee-Lucas Chair. R.F.O. was supported by an EPSRC Career Advancement Fellowship and Marie Curie IRG (PIRG08-GA-2010-277080).

## Author Contributions

M.P.N. and R.F.O. conceived of the experiments. M.P.N. developed the theoretical simulations, fabricated the samples, and conducted the experiments. X. S. prepared the MEH-PPV sample for Z-scan characterisation, which were conducted by M. P. N. and P. D. M.P.N. and R.F.O. co-wrote the manuscript. All authors commented on the manuscript.

## References

1. Li, Q., Davanço, M. & Srinivasan, K. Efficient and low-noise single-photon-level frequency conversion interfaces using silicon nanophotonics. *Nat. Photonics* **10,** 406–414 (2016).
2. Fras, F. *et al.* Multi-wave coherent control of a solid-state single emitter. *Nat. Photonics* **10,** 155–158 (2016).
3. Kachynski, A. V. *et al.* Photodynamic therapy by in situ nonlinear photon conversion. *Nat. Photonics* **8,** 455–461 (2014).
4. Ideguchi, T. *et al.* Coherent Raman spectro-imaging with laser frequency combs. *Nature* **502,** 355–358 (2013).
5. Harutyunyan, H., Beams, R. & Novotny, L. Controllable optical negative refraction and phase conjugation in graphite thin films. *Nat. Phys.* **9,** 423–425 (2013).
6. Zlatanovic, S. *et al.* Mid-infrared wavelength conversion in silicon waveguides using ultracompact telecom-band-derived pump source. *Nat. Photonics* **4,** 561–564 (2010).
7. Hugi, A., Villares, G., Blaser, S., Liu, H. C. & Faist, J. Mid-infrared frequency comb based on a quantum cascade laser. *Nature* **492,** 229–233 (2012).
8. Salem, R. *et al.* Signal regeneration using low-power four-wave mixing on silicon chip. *Nat. Photonics* **2,** 35–38 (2008).
9. Suchowski, H. *et al.* Phase mismatch-free nonlinear propagation in optical zero-index materials. *Science.* **342,** 1223–6 (2013).
10. Nielsen, M. P. *et al.* Adiabatic Nanofocusing in Hybrid Gap Plasmon Waveguides on the Silicon-on-Insulator Platform. *Nano Lett.* **16,** 1410–4 (2016).
11. Lafone, L., Sidiropoulos, T. P. H. & Oulton, R. F. Silicon-based metal-loaded plasmonic waveguides for low-loss nanofocusing. *Opt. Lett.* **39,** 4356–9 (2014).
12. Stockman, M. Nanofocusing of Optical Energy in Tapered Plasmonic Waveguides.

*Phys. Rev. Lett.* **93,** 137404 (2004).
13. Zhao, Y., Lombardo, D., Mathews, J. & Agha, I. All-optical switching via four-wave mixing Bragg scattering in a silicon platform. *APL Photonics* **2,** 26102 (2017).
14. Ho, M. C., Marhic, M. E., Wong, K. Y. K. & Kazovsky, L. G. Narrow-linewidth idler generation in fiber four-wave mixing and parametric amplification by dithering two pumps in opposition of phase. *J. Light. Technol.* **20,** 469–476 (2002).
15. Foster, M. a *et al.* Silicon-chip-based ultrafast optical oscilloscope. *Nature* **456,** 81–84 (2008).
16. Takesue, H. & Inoue, K. Generation of polarization-entangled photon pairs and violation of Bell's inequality using spontaneous four-wave mixing in a fiber loop. *Phys. Rev. A* **70,** (2004).
17. Harutyunyan, H., Volpe, G., Quidant, R. & Novotny, L. Enhancing the Nonlinear Optical Response Using Multifrequency Gold-Nanowire Antennas. *Phys. Rev. Lett.* **108,** 217403 (2012).
18. Palomba, S. *et al.* Optical negative refraction by four-wave mixing in thin metallic nanostructures. *Nat. Mater.* **11,** 34–38 (2011).
19. Renger, J., Quidant, R., van Hulst, N. & Novotny, L. Surface-Enhanced Nonlinear Four-Wave Mixing. *Phys. Rev. Lett.* **104,** 46803 (2010).
20. Kravtsov, V., Ulbricht, R., Atkin, J. M. & Raschke, M. B. Plasmonic nanofocused four-wave mixing for femtosecond near-field imaging. *Nat. Nanotechnol.* **11,** 459–464 (2016).
21. Aouani, H., Rahmani, M., Navarro-Cía, M. & Maier, S. A. Third-harmonic-upconversion enhancement from a single semiconductor nanoparticle coupled to a plasmonic antenna. *Nat. Nanotechnol.* **9,** 290–294 (2014).
22. Melikyan, A. *et al.* High-speed plasmonic phase modulators. *Nat. Photonics* **8,** 229–233 (2014).
23. Duffin, T. J. *et al.* Degenerate four-wave mixing in silicon hybrid plasmonic waveguides. *Opt. Lett.* **41,** 155–8 (2016).
24. Martin, S. J., Bradley, D. D. C., Lane, P. A., Mellor, H. & Burn, P. L. Linear and nonlinear optical properties of the conjugated polymers PPV and MEH-PPV. *Phys. Rev. B* **59,** 15133–15142 (1999).
25. An, L., Liu, H., Sun, Q., Huang, N. & Wang, Z. Wavelength conversion in highly nonlinear silicon–organic hybrid slot waveguides. *Appl. Opt.* **53,** 4886 (2014).
26. Koos, C. *et al.* All-optical high-speed signal processing with silicon–organic hybrid slot waveguides. *Nat. Photonics* **3,** 216–219 (2009).
27. Semple, J. *et al.* Radio Frequency Coplanar ZnO Schottky Nanodiodes Processed from Solution on Plastic Substrates. *Small* **12,** 1993–2000 (2016).
28. Afshar V., S. & Monro, T. M. A full vectorial model for pulse propagation in emerging waveguides with subwavelength structures part I: Kerr nonlinearity. *Opt. Express* **17,** 2298 (2009).
29. Savage, K. J. *et al.* Revealing the quantum regime in tunnelling plasmonics. *Nature* **491,** 574–577 (2012).
30. Nielsen, M. P., Ashfar, A., Cadien, K. & Elezzabi, A. Y. Plasmonic materials for metal-insulator-semiconductor-insulator-metal nanoplasmonic waveguides on silicon-on-insulator platform. *Optical Materials* (2013).

## Methods

### Sample Fabrication

The silicon hybrid gap plasmon waveguides presented here were fabricated with a two-step EBL method similar to that presented previously[10]. First 25 nm of $SiO_2$ was sputtered (Angstrom AMOD 520 resistive and sputter deposition system) onto a SOI substrate comprising of a 220 nm device layer and a 3 µm buried oxide layer. Then electron beam lithography (EBL, Raith eLine system) followed by 40 nm of Au evaporation and lift-off was used to define the coupling gratings and half of the waveguides. A second EBL step was used to define the second half of the waveguides in order to realise sub-100nm gaps. For the Au film adhesion, a self-assembled monolayer was used instead of a Cr or Ti adhesion layer to minimize propagation loss[30]. The inherent Au film roughness limits the resultant quoted gap widths to within ±5 nm. The MEH-PPV cladding layer was first created by dissolving 15.5 mg of poly[2-methoxy-5-(2-ethylhexyloxy)-1,4-phenylenevinylene] (MEH-PPV), number average molecular weight 70,000-100,000, from Sigma Aldrich in 1.6 mL of toluene overnight at 65°C. To coat the HGPW sample, the dissolved MEH-PPV was spin coated at 1000 RPM for 60 s to create a 120 nm film.

**Nonlinear Pulse Propagation Simulations**

The nonlinear pulse propagation simulations utilized the nonlinear Schrodinger equation as described in our previous publication[23]. The NLSE simulations consider co-propagating 1.04 ps pulses at 1450 nm and 1480 nm and account for dispersion up to the second order, phase matching, the Kerr effect, multiphoton absorption, and propagation losses, to correctly predict the spectrally broader idler spectrum centred at 1510 nm seen in Figure 2. The effective waveguide nonlinearity considered in the simulation was calculated using the measured third order nonlinear properties of Si, $SiO_2$, MEH-PPV, and Au from the Z-scan method (see Supplementary Information) and through mode solving simulations to account for the electromagnetic field overlap among the waveguide's constituent materials. For more information on the NLSE simulations conducted and how the effective waveguide nonlinearity varies with gap width, the reader is directed to the Supplementary information.

**Experimental Setup**

The experimental setup used in the degenerate four-wave mixing experiment is shown in the Supplementary Information. A 137 fs FWHM femtosecond pulses centred at 1480 nm at a 10 KHz repetition rate were produced by a Light Conversion Yb:KGW PHAROS system pumping an ORPHEUS optical parametric amplifier, and then fed into a 4F spatial filter. The 4F spatial filter was formed of two 600 lines/mm gratings with blaze wavelength $\lambda = 1600$ nm, two lens of 75 mm focal length, and an aperture with two slits in it. By varying the width and position of the two slits, two spectrally separated and temporally longer pulses (~1 ps) could be created for use as the pump and signal in the DFWM experiment. The input beams were then aligned to the input gratings of the sample using a 3-axis stage in a microscope setup. The emission from the gratings was spatially filtered with an aperture and focused into a LN-cooled 1-D InGaAs detector array in a Princeton Instruments OMA V SP2300 spectrometer with a 600 gr/mm at 1.6 µm wavelength blaze grating. When measuring the DFWM idler output, 1500 nm longpass filters were used to block the input pump/signal beams. Ambient light was kept to a minimum as MEH-PPV degrades when subject to light with above bandgap energies.